# Computational distributed fiber-optic sensing

**DA-PENG ZHOU,**[1,*] **WEI PENG,**[1] **LIANG CHEN,**[2] **AND XIAOYI BAO**[2]

[1]*School of Physics, Dalian University of Technology, Dalian, Liaoning 116024, China*
[2]*Department of Physics, University of Ottawa, Ottawa, ON K1N 6N5, Canada*
*\*dpzhou@dlut.edu.cn*

**Abstract:** Ghost imaging allows image reconstruction by correlation measurements between a light beam that interacts with the object without spatial resolution and a spatially resolved light beam that never interacts with the object. The two light beams are copies of each other. Its computational version removes the requirement of a spatially resolved detector when the light intensity pattern is pre-known. Here, we exploit the temporal analogue of computational ghost imaging, and demonstrate a computational distributed fiber-optic sensing technique. Temporal images containing spatially distributed scattering information used for sensing purposes are retrieved through correlating the "integrated" backscattered light and the pre-known binary patterns. The sampling rate required for our technique is inversely proportional to the total time duration of a binary sequence, so that it can be significantly reduced compared to that of the traditional methods. Our experiments demonstrate a 3 orders of magnitude reduction in the sampling rate, offering great simplification and cost reduction in the distributed fiber-optic sensors.

## 1. Introduction

Conventional time-domain techniques used by distributed fiber-optic sensors (DFOSs) rely on sending an optical pulse into an optical fiber, and collecting backscattered light by a photodetector [1-5]; the resultant time-domain electrical signal needs to be acquired by a fast-speed digitizer whose sampling rate is required to meet the Nyquist-Shannon sampling theorem [6], so that the sampling time interval must be shorter than half of the pulse duration which determines the spatial resolution of the sensors [1]. DFOSs also require data acquisition systems to have high resolution to provide large dynamic range in order to achieve long range measurement. Note that this continuous long-range monitoring capability is a superior advantage of the DFOSs compared to their discrete counterparts [1-5]. Hence, fast transferring and processing large amount of data with high resolution demands sophisticated designing and programming. This in turn keeps the DFOSs at generally higher prices than conventional sensors, which nowadays is still one of the major factors preventing DFOSs for wider applications in practice.

The requirement for high sampling rate in the time domain is by analogy with the necessity of sufficient spatial resolution for cameras used in a conventional optical imaging scheme. However, in recent years, the need for spatially resolved detectors can be removed thanks to the single-pixel camera (SPC) techniques which are closely related to computational ghost imaging (CGI) in terms of the experiment arrangements [7-10]. Classical ghost imaging (GI) [11,12] is a novel technique performing correlation measurements between light transmitted through or reflected from an object collected by a "bucket" photodetector without spatial resolution and the spatially resolved intensity pattern of the incident light measured by a detector array to retrieve the ghost image of the original object [13,14]. When the object is illuminated by a light beam with pre-known intensity patterns, the resultant scheme is called CGI or SPC so that the spatially resolved detector are no longer needed [7-10]. Recently, GI technique has been extended to the time domain by exploiting space-time duality of optics [15], with a classical nonstationary light source [16,17], a chaotic laser [18], biphoton states [19], or a quasi-continuous multimode laser source [20] to produce random illuminations. In [20], a version of the GI protocol creates "temporal" instead of "spatial" images, retrieving an embedded binary signal with a good signal-to-noise ratio (SNR), offering great potential for dynamic imaging of ultrafast waveforms [21]. Later on, thermal temporal GI [22] is demonstrated, and temporal CGI is also proposed to retrieve a non-reproducible single temporal signal [23] and to detect fast signals beyond the bandwidth of the detectors [24].

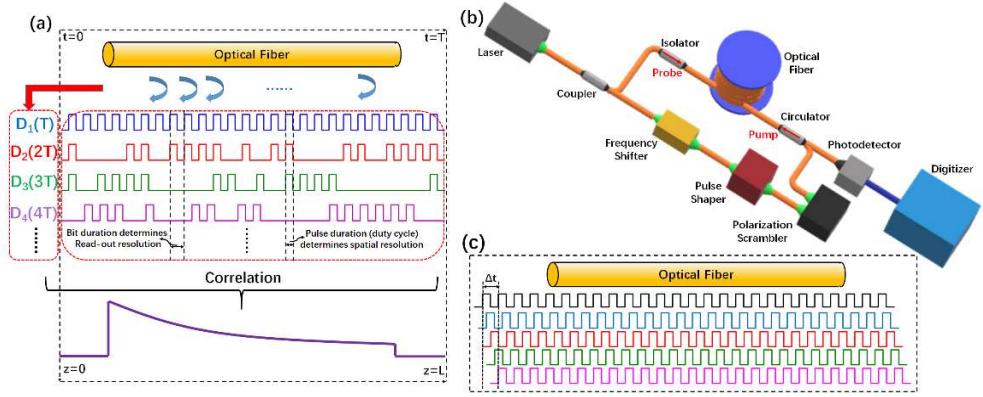

Fig. 1. Schematic of the proposed computational distributed fiber-optic sensing system. (a) Binary pulse pattern sequences are sent into an optical fiber, and the backscattered integrated signals are received by a photodetector followed by a digitizer. Each pulse sequence has a total duration of $T$, so that the sampling rate of the digitizer is $1/T$. Correlation calculations are performed between the pre-known pulse pattern and the acquired integrated signal retrieving the temporal image containing scattering information. (b) Experiment setup utilizing stimulated Brillouin scattering to demonstrate the proposed sensing technique. (c) Illustration of trigger delaying method to increase the read-out spatial samples. The reconstruction process is repeated by 5 times, and each time the trigger is delayed by $\Delta t/5$ with $\Delta t$ being bit duration of the binary pulse sequence, so that each time the reconstructed temporal image is shifted by $\Delta t/5$. After interleaving the 5 images, one can obtain the entire temporal image with more detailed location information.

Inspired by GI techniques, we demonstrate a computational distributed fiber-optic sensing technique with a sampling rate far below the Nyquist rate which is complied with most of the conventional time-domain DFOSs. Unlike the applications to retrieve ultrafast active temporal objects [20,21,23,24], our proposed technique opens new perspectives based on GI to retrieve passive stationary temporal objects which is very suitable for DFOSs as many of them are developed for non-dynamic sensing environment. In fact, the spatially distributed scattering information along a sensing fiber can be "viewed" as a spatial "image". Physically speaking, since DFOSs measure this distributed information by sending an optical pulse and detecting its backscattering along the fiber, the spatial information is then carried by the backscattered optical signals in the time domain; therefore, the spatial "image" is transformed into a temporal object or a temporal "image". It is feasible to obtain this temporal "image" without the necessity of continuous fast-speed acquisitions. We send optical pulse sequences with pre-known binary patterns to illuminate the spatial scattering object sequentially in time domain, and in the meantime, the sensing fiber itself "integrates" backscattered light continuously and the resultant integrated optical signal only needs to be collected at discrete specific moments with an acquisition rate inversely proportional to the total time duration of the light sequence as illustrated in Fig. 1(a). Correlation calculations between the collected data and the pre-known binary patterns are performed afterwards to retrieve the temporal "image". Note that this process is different from the temporal GI [20,21,24] using a slow detector as the detector integrates optical signals in electrical domain, while the proposed computational DFOS integrates optical signals in optical domain. For sensing systems based on Rayleigh scattering, spontaneous Raman scattering and spontaneous Brillouin scattering, the sensing fibers can serve as good optical integrators in the backward direction. For sensing systems based on stimulated Brillouin scattering, temporal GI can also be adopted when the powers of two counter-propagating light beams are adjusted to the levels that pump depletion and excess amplification can be ignored. This condition is also the requirement for high performance Brillouin sensing system [25]. In this work, we use stimulated Brillouin scattering as the sensing mechanism to demonstrate the computational distributed fiber-optic sensing technique, primarily because Brillouin frequencies varies along optical fiber which results in a number of different frequency-dependent temporal images in only one experimental setup, providing solid evidences that the proposed method can be applied to many other kinds of time-domain DFOSs. Our technique can achieve a sampling rate much lower than traditional methods, and more than 3 orders of magnitude of reduction can be realized experimentally.

## 2. Operation principle and experiment setup

*2.1 Operation principle*

The proposed computational distributed fiber-optic sensing technique is based on computational differential GI (DGI) technique in the time domain. Conventional GI protocol using random binary patterns requires, normally, a large number of iterations to reconstruct the image [26,27], and results in low SNR especially when scattering coefficients are very small. We use DGI protocol to enhance the SNR [26,27] as it utilizes a weighting factor that depends on fluctuations in the measured signal and facilitates image retrieving in absolute unit. In this work, we first realize computational distributed fiber-optic sensing using random sequences. However, decent SNR of the DGI using random patterns still requires relatively large number of iterations, since the SNR is proportional to the number of iterations [26]. In order to further reduce the number of measurements, we later use Walsh-Hadamard sequences instead of random sequences. Since Walsh-Hadamard patterns are orthogonal [28,29], we will show in this section that Walsh-Hadamard pattern pairs based DGI protocol is equivalent to the process of an inverse Walsh-Hadamard transform of the acquired differential signal that makes the reconstruction exact if no measurement noise is present. In our experiment, we intend to deal with small number of time-domain "pixels"; therefore, Walsh-Hadamard binary sequences can further reduce the number of measurements significantly compared with the random sequences, which paves the way to real applications of our proposed technique.

The light is modulated by a semiconductor optical amplifier to produce pairs of Walsh-Hadamard patterns as shown in Fig. 1(a), which are generated from each row of a natural-order Walsh-Hadamard matrix with a dimension of $2^k$,

$$\boldsymbol{H}_{2^k} = \begin{bmatrix} \boldsymbol{H}_{2^{k-1}} & \boldsymbol{H}_{2^{k-1}} \\ \boldsymbol{H}_{2^{k-1}} & -\boldsymbol{H}_{2^{k-1}} \end{bmatrix}, \qquad (1)$$

which is the recursive formula for $k \geq 2$ from the lowest order matrix

$$\boldsymbol{H}_{2^1} = \begin{bmatrix} 1 & 1 \\ 1 & -1 \end{bmatrix}. \qquad (2)$$

We produce Walsh-Hadamard pattern $I_i(t)$ and its inverse pattern $\tilde{I}_i(t)$ from each row of the matrix, where $i = 1, 2 \cdots 2^k$ with the integer $k$ determining the length of the Walsh-Hadamard sequence, and $t = t_1, t_2 \cdots t_{2^k}$ represents discrete time sequence corresponding to each binary bit. The values of $I_i(t_i)$ is binary, either 1 for the value +1 in $\boldsymbol{H}_{2^k}$ or 0 for the value -1 in $\boldsymbol{H}_{2^k}$, and $\tilde{I}_i(t) = 1 - I_i(t)$. The total duration of a sequence becomes $T = 2^k \Delta t$ with $\Delta t$ being a bit period; therefore, the sampling rate of the backscattered signal is $1/T$ which is much lower than that the conventional method needs. The detected signals $D_i$ using Walsh-Hadamard pattern and $\widetilde{D}_i$ using the inverse pattern can be expressed as

$$D_i = \gamma \int_0^T I_i(t) S(t) \mathrm{d}t, \qquad (3)$$

$$\widetilde{D}_i = \gamma \int_0^T \tilde{I}_i(t) S(t) \mathrm{d}t, \qquad (4)$$

where $\gamma$ takes account the responsivity and amplifier gain of the photodetector. $S(t)$ is the amount of power transferred between the pump power $P_P(z)$ and the probe $P_s(z)$ at location $z = ct/2n$ due to stimulated Brillouin scattering at a fixed frequency difference between pump and probe, where $n = 1.4682$ which is the group refractive index of the fiber around 1550 nm, and $c$ is the velocity of light in vacuum. It contains distributed scattering properties and can be treated as the temporal image to be reconstructed [25]:

$$S(t) = S(z) \approx \frac{g(z)}{A_{eff}} P_P(z) P_s(z) \Delta z, \tag{5}$$

where $g(z)$ is the local Brillouin gain coefficient, $A_{eff}$ is the effective mode area, and $\Delta z = c\Delta t/2n$ is the interaction length. Note that Eq. (5) is under the condition that pump depletion and excess amplification is avoided by properly adjusting pump and probe powers at both ends of the fiber, because high performance Brillouin optical time-domain analysis (BOTDA) sensors are required to operate in a small gain region and the relative power transfer is very small [25]. For Rayleigh, spontaneous Raman or Brillouin scattering, $S(t)$ becomes the amount of scattered power in the backward direction without the continuous-wave probe, and can be expressed as

$$S(t) = S(z) \approx \alpha(z) P_P(z) \Delta z, \tag{6}$$

with $\alpha(z)$ being the corresponding local backscattering coefficient. Note that in this case, the experiment arrangement should be adjusted correspondingly with respect to different scattering mechanisms. Note that in Eqs. (5) and (6), both $g(z)$ and $\alpha(z)$ are assumed constant over $\Delta z$, and the quantity $S(t)$ corresponds to the local response containing sensing information.

Therefore, the temporal image $S(t)$ calculated over a series of time-domain measurement using DGI protocol is defined by [26] as illustrated in Fig. 1(a),

$$\gamma S(t) = \frac{<D>}{<R>} + \frac{1}{<I(t)>^2} [<DI(t)> - \frac{<D>}{<R>} <RI(t)>], \tag{7}$$

where $<\cdots>$ denotes an ensemble average, and

$$R_i = \int_0^T I_i(t) dt, \tag{8}$$

$$\tilde{R}_i = \int_0^T \tilde{I}_i(t) dt, \tag{9}$$

are the references which are pre-known and do not need to be measured. In Eq. (7), all of the subscripts $i$ have been removed from the notations to illustrate that the averages involve both Walsh-Hadamard patterns and its inverse patterns. Note that for the case using random patterns in Eq. (7), the SNR is proportional to the number of iterations [26], and it usually requires large number of averages to achieve a decent SNR even without considering detector noise. However, for the case that uses Walsh-Hadamard patterns which are orthogonal, the temporal image reconstruction time is significantly reduced especially for the case of small $k$ number, because it requires $N = 2^{k+1}$ averages in Eq. (7), and the process is equivalent to perform an inverse Walsh-Hadamard transform of the differential signals $D_i - \tilde{D}_i$ which makes the reconstruction exact. By using the properties of the Walsh-Hadamard matrix, and after some straightforward algebra, Eq. (7) can be reduced to a matrix form with the help of Eqs. (3), (4), (8) and (9),

$$\gamma \begin{bmatrix} S(t_1) \\ S(t_2) \\ \vdots \\ S(t_{2^k}) \end{bmatrix} = H_{2^k}^{-1} D = \frac{1}{2^k} H_{2^k} D =$$

$$\frac{1}{2^k} \begin{bmatrix} I_1(t_1) - \tilde{I}_1(t_1) & I_1(t_2) - \tilde{I}_1(t_2) & \cdots & I_1(t_{2^k}) - \tilde{I}_1(t_{2^k}) \\ I_2(t_1) - \tilde{I}_2(t_1) & I_2(t_2) - \tilde{I}_2(t_2) & & I_2(t_{2^k}) - \tilde{I}_2(t_{2^k}) \\ \vdots & & \ddots & \vdots \\ I_{2^k}(t_1) - \tilde{I}_{2^k}(t_1) & I_{2^k}(t_2) - \tilde{I}_{2^k}(t_2) & \cdots & I_{2^k}(t_{2^k}) - \tilde{I}_{2^k}(t_{2^k}) \end{bmatrix} \begin{bmatrix} D_1 - \tilde{D}_1 \\ D_2 - \tilde{D}_2 \\ \vdots \\ D_{2^k} - \tilde{D}_{2^k} \end{bmatrix} \tag{10}$$

where $\boldsymbol{D}$ represents the acquired differential signal vector. Therefore, the entire sensing process of the computational DGI using Walsh-Hadamard pattern pairs can be viewed as measuring Walsh-Hadamard coefficients $D_i - \widetilde{D}_i$ of the sensing information first, and then perform an inverse transform. In principle, it can provide exact "image" reconstruction if there are no noises in the system, because Walsh-Hadamard matrix is orthogonal without overlap information and redundancy. It is worth mentioning that the required number of measurements is twice compared to the number of time-domain "pixels", so when the number of "pixels" is large, one still needs a large number of iterations. Hence, for the case that a decent SNR is acceptable with large number of "pixels", using random sequences may be good enough and require less iterations, especially when compressive sampling [6,30] technique is used. However, for our case to realize computational distributed fiber-optic sensing towards practical applications, we intend to deal with small number of "pixels", so that a significant reduction in the number of measurements can be achieved compared to the case that random sequences are used which will be demonstrated experimentally in the later sections. Moreover, in a practical sensing system after detecting $D_i$ and $\widetilde{D}_i$, one can use either Eq. (7) or Eq. (10) for image reconstructions.

*2.2 Experiment setup*

The schematic diagram of the experiment setup is shown in Fig. 1(b). A laser beam with a wavelength of 1554 nm from a semiconductor laser is split by an optical coupler; one path goes directly to the optical fiber under test through an optical isolator forming a continuous-wave probe and the other path is frequency shifted by a carrier-suppressed optical single sideband modulator and then the power is boosted by an erbium-doped fiber amplifier (We refer both the modules together as a frequency shifter in Fig 1(b)). The resultant beam is modulated into optical pulses by a pulse shaper which in our experiment is a semiconductor optical amplifier driven by a function generator. The pulses referred as pump are polarization scrambled and then delivered into the optical fiber through an optical circulator. The signal out of the photodetector with a bandwidth of 45 MHz is acquired by a 14-bit high-resolution digitizer and the results are analyzed on a computer. Both the probe power and pump peak power can be adjusted independently to assure small gain criteria.

## 3. Experiment results

*3.1 Distributed sensing for short distance*

Our first goal is to measure distributed Brillouin spectra along a 1 km optical fiber by reconstructing time-domain traces of different frequency components around the Brillouin frequency of the fiber. The 1 km fiber has a Brillouin frequency around 10860 MHz. Close to the end of the fiber, we directly splice a section of 20 m fiber with a Brillouin frequency around 10790 MHz. The introduction of this fiber section is to simulate an event of temperature change or strain variation since Brillouin sensors are measuring distributed Brillouin frequency variations which reflect the external perturbations. This type of fiber arrangement also provides spatial resolution information by examining the transition region on the time-domain traces at the splicing point because of the large sudden Brillouin frequency change.

We adopt DGI protocol [26,27] with a Walsh-Hadamard matrix of dimension $2^8$ resulting in $N = 2^9$ different pulse pattern sequences which are sent into the fiber for retrieving temporal image at each frequency component. Each optical sequence uses return-to-zero unipolar encoding format with 50% duty cycle to avoid pump depletion and excess amplification. The duration of a single bit is 50 ns, which corresponds to a 5 m read-out resolution. However, the actual physical spatial resolution is 2.5 m which is determined by effective pulse duration of 25 ns. Thus, the total duration of one sequence is 12.8 μs, which determines the acquisition sampling rate of 78.125 kHz. Figs. 2(a) and 2(b) show reconstruction results obtained by DGI protocol using Walsh-Hadamard patterns (WHGI) for two different frequencies of 10860 MHz and 10790 MHz, respectively (see Visualization 1 and Visualization 2 for a full evolution of the temporal image as a function of the number of realizations). As expected, the results coincide with those obtained by inverse Walsh-Hadamard transform (IWHT) of the acquired differential signal. We also show the results obtained by DGI technique using random binary pattern sequences (RSGI), where 4096 iterations are carried out. It can be seen clearly that the SNR of WHGI is much better than that of the RSGI, even though the latter method uses much more iterations, because the Walsh-Hadamard patterns are orthogonal without any redundancy [28,29]. We also compare all the reconstructed results with the conventional method by sending a 25 ns pulse into the fiber and acquiring the return signal with the same read-out resolution for the comparison purpose. The results agree well with each other, verifying the accuracy of the proposed method.

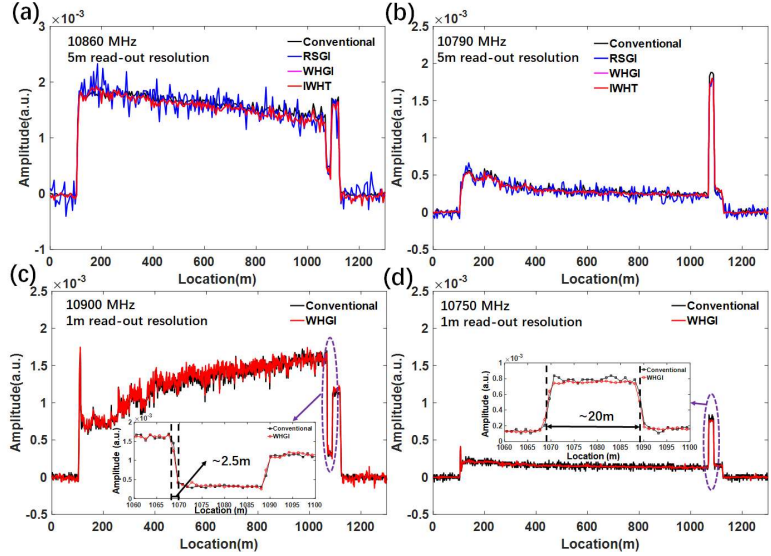

Fig. 2. Reconstructed temporal images at several different frequencies for 1 km fiber. (a) and (b) show the results at frequencies of 10860 MHz and 10790 MHz, respectively. The spatial distance between two neighboring points is 5 m determined by the bit period of 50 ns. The WHGI reconstructed images are also compared with the time-domain traces obtained by RSGI, IWHT and the conventional method using a single 25 ns pulse with 5 m read-out resolution. (c) and (d) show the results at frequencies of 10900 MHz and 10750 MHz, respectively. The spatial distance between two neighboring points is 1 m by interleaving 5 reconstructed temporal images of 5 m read-out resolution. The WHGI reconstructed images are also compared with the time-domain traces obtained by conventional method using a single 25 ns pulse with 1 m read-out resolution. The insets show enlarged view of the 20 m fiber confirming the 2.5 m spatial resolution.

Next, we increase the read-out resolution to 1 m in order to verify the 2.5 m spatial resolution. A straightforward way is to repeat the reconstruction procedure for 5 times. At each time, the trigger is delayed by 10 ns before receiving by the digitizer, so that the reconstructed temporal image shifts 10 ns (corresponding to 1 m shift spatially) each time as illustrated in Fig. 1(c). In other words, each shifted image can be considered as additional samples from different locations of the sensing fiber. Therefore, the resultant 5 WHGI results can be interleaved to form a complete temporal image with 1 m spatial read-out resolution. Please note, the spatial resolution is not increased as it is determined by the effective pulse duration due to the 50% duty cycle. However, for demonstration purposes, we directly acquire 5 points at every 10 ns, and the results in principle should be the same as the trigger delaying method mentioned above. The resultant interleaved temporal images with 1 m read-out resolution is shown in Figs. 2(c) and 2(d) for two different frequencies of 10900 MHz and 10750 MHz, respectively. At the end of the fiber, the 20 m fiber is clearly indicated. The leading and trailing edges of the traces approximately equal to 2.5 m as expected. The results are compared with conventional method using a single 25 ns pulse with 100 MHz sampling rate, and the two results agree with each other very well. It is worth mentioning that, in practical applications, one can use 5 analog-to-digital converters (ADCs) simultaneously to acquire the data at the same repetition rate of 78.125 kHz, but adjusting the trigger delay for each ADC to complete the process at once. Thus, for this example, the WHGI technique reduces the sampling rate by more than 3 orders of magnitude with respect to the conventional method. Note that even though there are fast algorithms for IWHT, WHGI method can reconstruct the image during the acquisition process; therefore, we only select WHGI technique throughout the rest of the paper. Next, we scan the frequency difference between pump and probe from 10500 MHz to 11200 MHz at 1 MHz increment to obtain Brillouin spectra along the optical fiber. Fig. 3(a) shows the results using conventional method with a single 25 ns pulse, while Fig. 3(b) shows the results using WHGI method. Both the figs. have 1 m read-out spatial step. The distributed Brillouin frequencies are obtained by Lorentzian fitting as shown in Fig. 3(c). The 20 m fiber section with different Brillouin frequency at the end of the fiber is seen clearly, verifying the effectiveness of the proposed method to measure temperature or strain variations along the fiber.

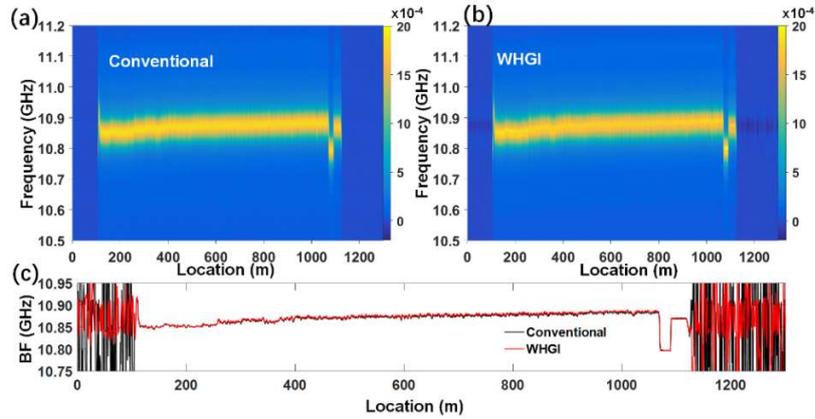

Fig. 3. Comparison of the results obtained by conventional method and the proposed WHGI method for the 1 km fiber. Brillouin spectra obtained by (a) conventional method using a single 25 ns pulse and (b) the proposed WHGI method. The read-out resolution for both the figs. is 1 m, and the frequency is scanned from 10500 MHz to 11200 MHz with 1 MHz increment. (c) the distributed Brillouin frequency (BF) using the two methods along the 1 km fiber by Lorentzian fitting of the Brillouin spectra in (a) and (b).

*3.2 Distributed sensing for long distance*

The second example is to achieve long range measurement, as this ability is one of the superior advantages of the DFOSs. We test 51 km fibers by connecting three fiber spools: the leading 2 fiber spools with 25 km each, and the third fiber spool is the same 1 km fiber used in Figs. 2 and 3. We choose 100 ns bit duration which corresponds to 10 m read-out resolution, so that the actual spatial resolution is 5 m because of the 50% duty cycle in a bit period. Based on the measurement length, a straightforward method is to use the Walsh-Hadamard sequence long enough to cover the entire fiber length. However, for 51 km fiber, $2^{13}$ bits are required resulting in a considerably long acquisition time for reconstruction. Instead, shorter Walsh-Hadamard pattern sequence can be used to cover a partial section of fiber, so that the entire temporal image can be obtained after each section are reconstructed. The sampling rate for the digitizer is still determined by the duration of a Walsh-Hadamard sequence. We select $2^7$ bits corresponding to 12.8 μs sequence, which determines the sampling rate of 78.125 kHz; therefore, 40 sections can cover over 51 km measurement range.

The results are shown in Figs. 4(a) and 4(b) for two different frequencies of 10860 MHz and 10790 MHz, respectively (see Visualization 3 and Visualization 4 for a full evolution of the temporal image as a function of the number of sections). All the results shown in the figs. have 2 m spatial step after interleaving 5 independent temporal images with a 20 ns shift each. The results using conventional method by sending a single 50 ns pulse into the fiber and acquiring with 50 MHz sampling rate are also shown in the figs. for comparison. Both results agree well with each other, and all the three sections of the fiber spools are clearly shown. At the end of the fiber, the 20 m testing fiber is observable (see insets in the figs.), confirming that 5 m spatial resolution is achieved as expected. Finally, we scan the frequency difference between pump and probe from 10600 MHz to 11200 MHz at 1 MHz increment to obtain Brillouin spectra along the 51 km optical fiber. Figs. 5(a) and 5(b) show the results using conventional and WHGI method, respectively. Both the figs. have 2 m read-out spatial step. The distributed Brillouin frequencies are shown in Fig. 5(c), and the 20 m fiber section at the end of the fiber is resolved clearly.

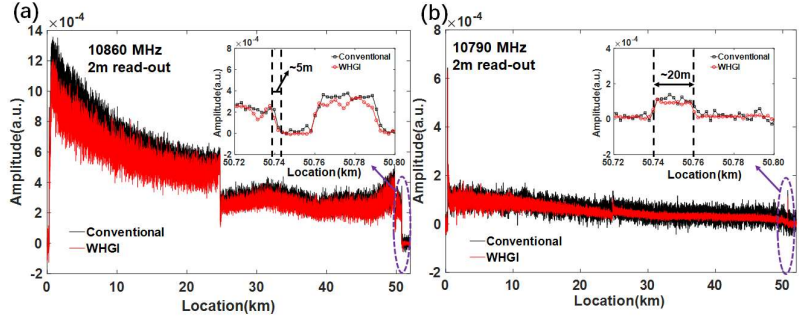

Fig. 4. Reconstructed temporal images at two different frequencies for 51 km fiber. (a) and (b) show the results at frequencies of 10860 MHz and 10790 MHz, respectively. The spatial distance between two neighboring points is 2 m by interleaving 5 reconstructed temporal images of 10 m read-out resolution. The WHGI reconstructed images are also compared with the time-domain traces obtained by conventional method using a single 50 ns pulse with 2 m read-out resolution. The insets show enlarged view of the 20 m fiber confirming the 5 m spatial resolution.

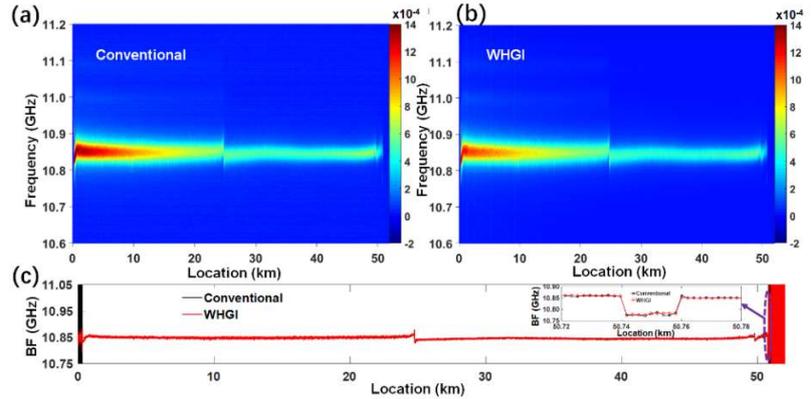

Fig. 5. Comparison of the results obtained by conventional method and the proposed WHGI method for the 51 km fiber. Brillouin spectra obtained by (a) conventional method using a single 50 ns pulse, and (b) the proposed WHGI method. The read-out resolution for both the figs. is 2 m, and the frequency is scanned from 10600 MHz to 11200 MHz with 1 MHz increment. (c) the distributed Brillouin frequency (BF) using the two methods along the 51 km fiber by Lorentzian fitting of the Brillouin spectra in (a) and (b).

## 4. Discussion

The sampling rate of the proposed technique is inversely proportional to the total time duration of a pulse sequence; therefore, it can be reduced significantly without compromising the acquisition accuracy. Either random binary sequences or Walsh-Hadamard binary sequences can be used in our technique. We demonstrate experimentally that the sampling rate can be reduced by over 3 orders of magnitude compared to the conventional technique. Please note that only the sampling rate can be reduced, but the bandwidth of the detection circuits in the system should be the same as that of conventional methods, since in our method the integration is performed in optical domain. The sampling rate can be set lower by using longer sequence; however, it will take more averages due to the symmetric property of the Walsh-

Hadamard matrix. Hence, there is a tradeoff between the sampling rate and the total number of measurements. For long distance measurement as demonstrated in our experiment, the pulse sequence is not necessary to cover the entire sensing fiber. Once a suitable sampling rate is determined, the entire measurement range can be separated into sections that can be retrieved individually. The amount of collected data is two times more than that obtained from the conventional measurements, because both the Walsh-Hadamard sequence and the inverse sequence are used. Since Walsh-Hadamard sequence can be generated from a well-defined matrix form, only the acquired data needs to be stored and the time-domain traces can be reconstructed whenever it is needed; therefore, the assigned memory space to store the raw data is reduced significantly compared to that using random sequences.

Note that even though the present method implements Walsh-Hadamard pattern sequence for temporal image reconstruction, its operation principle is fundamentally different from the conventional method utilizing coding technique which is developed to enhance the system SNR due to the code gain [3,4]. Conventional coding technique still needs to acquire the entire time-domain traces with high sampling rate and processes them afterwards based on known mathematical properties of the selected code. Our method adopts GI technique which can be realized using classically correlated light beams driven by statistically independent noise processes [20]. We demonstrate that random sequences can be used to reconstruct the temporal images; however, it requires relatively large amount of iterations to complete reconstruction with decent SNR [26,27] especially for small number of time-domain "pixels". Hence, we take one step further by using DGI protocol with Walsh-Hadamard pattern pairs, and show that it is equivalent to perform an inverse Walsh-Hadamard transform of the differential acquired data to reduce the number of measurements.

## 5. Conclusion

In conclusion, the proposed technique employs computation DGI protocol with either random binary sequences or Walsh-Hadamard pattern binary sequences realizing computational distributed fiber-optic sensing. The operation principle is very different from conventional time-domain technique that requires continuously data acquisition governed by the Nyquist-Shannon sampling theorem. The acquisition sampling rate of our proposed technique is inversely proportional to the total time duration of a binary sequences, so it can be reduced significantly compared to conventional method, and over 3 orders of magnitude reduction is experimentally demonstrated in this work. The parameters demonstrated in the paper is sufficient for many practical applications. However, for certain applications which require higher spatial resolution, differential pulse-width pair technique [31] can be adopted easily because spatial resolution of the proposed method is still determined by the pulse duration as that of the conventional technique. Our results open great perspectives in designing acquisition systems for DFOSs with much less complexity and cost, and can be applied to a variety of time-domain DFOSs for long-range non-dynamic measurements including Rayleigh-scattering-based optical time-domain reflectometry, Raman-scattering-based distributed temperature sensors, and Brillouin optical time-domain analysis and reflectometry.

## Funding

Fundamental Research Funds for the Central Universities (DUT17RC(3)074); National Natural Science Foundation of China (61727816, 61520106013).

## Acknowledgments

D. –P. Zhou would also like to thank the financial support of the Academic Program Development Funds from Dalian University of Technology.